\documentclass{aa}  

\usepackage{graphicx}
\usepackage{txfonts}
\usepackage[caption = false]{subfig}
\usepackage{appendix}
\usepackage{natbib}
\bibpunct{(}{)}{;}{a}{}{,}

\begin{document}

   \title{Analysis of the SFR - M$^{*}$ plane at z$<$3: single fitting versus multi-Gaussian decomposition}

   \subtitle{}

   \author{L. Bisigello
          \inst{1,2}
          , K. I. Caputi\inst{1} , N. Grogin \inst{3}, A. Koekemoer\inst{3}
          }

   \institute{Kapteyn Astronomical Institute, University of Groningen,
              9747 AD, Groningen, The Netherlands\\
         \and
             SRON Space Research of Netherlands, 9747 AD, Groningen, The Netherlands\\
         \and
                Space Telescope Science Institute, 3700 San Martin Drive, Baltimore, MD 21218, USA\\
             }

   \date{Received ; accepted }
\titlerunning{Analysis of the SFR - M$^{*}$ plane at z$<$3: single fitting versus multi-Gaussian decomposition}
\authorrunning{Bisigello et al.}
 
  \abstract
  {The analysis of galaxies on the star formation rate - stellar mass (SFR-$\rm M^\ast$) plane is a powerful diagnostic for galaxy evolution at different cosmic times. We consider
a sample of 24463 galaxies from the CANDELS/GOODS-S survey to conduct a detailed analysis of the SFR-$\rm M^\ast$ relation at redshifts 0.5$\leqslant z<$3 over more than three dex in stellar mass. To obtain SFR estimates, we utilise mid- and far-IR photometry when available, and rest-UV fluxes for all the other galaxies. We perform our analysis in different redshift bins, with two different methods: 1) a linear regression fitting of all star-forming galaxies, defined as those with specific star formation rates $\rm log_{10}(sSFR/yr^{-1}) > -9.8$, similarly to what is typically done in the literature; 2) a multi-Gaussian decomposition to identify the galaxy main sequence (MS), the starburst sequence and the quenched galaxy cloud. We find that the MS slope becomes flatter when higher stellar mass cuts are adopted, and that the apparent slope change observed at high masses depends on the SFR estimation method. In addition, the multi-Gaussian decomposition reveals the presence of a starburst population which increases towards low stellar masses and high redshifts. We find that starbursts make up $\sim5\%$ of all galaxies at $z=0.5-1.0$, while they account for $\sim 16 \%$ of galaxies at $2<z<3$ with log$_{10}(M^{*}/M_{o})=$8.25-11.25. We conclude that the dissection of the SFR-$\rm M^\ast$ in multiple components over a wide range of stellar masses is necessary to understand the importance of the different modes of star formation through cosmic time. }



   \maketitle
%

\section{Introduction}
The majority of star-forming (SF) galaxies follow a relatively tight relation between stellar mass (M$^{*}$) and star formation rate (SFR) \citep{Noeske2007,Brinchmann2004}, the so-called star-forming main sequence (MS), up to high redshifts \citep{Whitaker2012,Steinhardt2014,Tasca2015}. This relation is key to understanding the evolution of galaxies and the interplay between secular effects, like gas depletion and gas infall, and more stochastic effects, such as mergers \citep{Peng2014,Tacchella2016}. For this reason, in the last decade, several works have been devoted to studying this relation and its evolution with redshift \citep{ Oliver2010,Karim2011,Rodighiero2011,Whitaker2014,Johnston2015,Shivaei2015,Kurczynski2016}.\par

In spite of this wide range of studies, a general consensus on the MS slope and its evolution with redshift has not been reached. This is because different sample selections, SFR derivations, and fitting techniques influence the derived MS making the comparison between different works difficult. For example, as explained by \cite{Speagle2014}, the MS derived using criteria that select blue, highly star-forming and not-dusty galaxies will be steeper than the MS derived with other selection mechanisms.  Ideally, having no pre-selection of the star-forming population, by using colours or a specific star formation rate (sSFR) cut, would seem the right approach to avoid influencing the derived MS slope. However, a sample with no pre-selection of SF galaxies could be contaminated by quenched galaxies, particularly at high stellar masses, producing a flatter MS than that derived from samples containing only SF galaxies.  \par

To have a MS that does not require pre-selection and is not contaminated by quenched galaxies, \citet{Renzini2015} defined the MS as the ridge line in the SFR-stellar mass-number density three-dimensional (3D) distribution surface. With this definition they could clearly distinguish between SF galaxies in the MS and quenched galaxies below the MS at redshift z$\sim$0.05, without influencing the slope of the derived MS. \par

As discussed exhaustively by \cite{Calzetti2013}, different SFR indicators are used across the wavelength spectrum, including X-ray, ultra-violet (UV), optical, infrared (IR), radio, and emission lines, based on direct stellar light, dust-processed light or ionised gas emission. SFR derived from the UV light has been used many times because it is dominated by the light of very young stars, and therefore is optimal to trace recent SF. Moreover, it is possible to derive the SFR in a wide range of redshifts observing the rest-frame UV redshifted to optical wavelengths. However, the presence of dust obscuring the UV light makes the derived SFR uncertain.\par

After the launches of  the {\em Spitzer Space Telescope} and \textit{Herschel Space Telescope}, observations at IR wavelengths started to be used more commonly to derive SFR, particularly for dusty galaxies. However, it is not possible to obtain IR data for large galaxy samples, and IR observations are less sensitive and have a smaller angular resolution than UV-optical observations. Moreover, several observations in the IR are necessary to reconstruct the IR spectral energy distribution (SED) at 3-1000 $\mu$m and therefore L$_{IR}$, which is then linked to the SFR. A series of relations between monochromatic IR luminosities and total IR luminosities have been derived during recent years \citep[e.g.][]{Chary2001,Bavouzet2008,Rieke2009}. However, these relations were obtained from different samples, at different redshift ranges, and using different monochromatic IR luminosities, and therefore the derived MS could be influenced by the relation used to convert from monochromatic IR luminosities and L$_{IR}$.\par

The aim of this work is to analyse the $SFR-M^{*}$ plane from z$=$0.5 to z$=$3, by comparing the traditional MS derivation method, after pre-selecting the SF galaxies with sSFR cut, with the ridge line definition from \citealt{Renzini2015}, obtained by fitting three Gaussian components to the sSFR distribution. In addition, we also analysed the effect of different SFR$_{IR}$ derivation methods on the bright-end of the MS. \par

The structure of the paper is as follows. In Section \ref{sec:sample} we present our sample selection, while in Section \ref{sec:zphot_mass} we describe the derivation of photometric redshifts and stellar masses by SED fitting. In Section \ref{sec:SFR} we report the different methods used to derive SFR from the UV and from the IR. In Section \ref{sec:MSfit} we analyse the sources on the $SFR-M^{*}$ plane, while in Section \ref{sec:summary} we summarise our main findings and conclusions. 
Throughout this paper, we adopt an H$_0=$70 kms$^{-1}$Mpc$^{-1}$, $\Omega_M=$0.27, $\Omega_\Lambda=$0.73 cosmology and a \cite{Salpeter1955} initial mass function over (0.1-100) M$_{\odot}$.

\section{Sample selection}\label{sec:sample}
\subsection{Data}
The GOODS-S field (\citealt{giavalisco+2004}) has been targeted with deep observations at different wavelengths over more than a decade. It is centred at $\alpha$(J2000)=$3^{h} 32^{m} 20^{s}$ and $\delta$(J2000)=$-27^{\circ}48^{\prime\prime}20^{s}$ and covers an area of approximately $10^{\prime}\times16^{\prime}$. In this study, we collect datasets in this field from the UV to the mid-IR, from both ground-based and space telescopes. \par

In particular, in the optical and near-IR, the GOODS-S field has been observed with the \textit{Hubble Space Telescope} (\textit{HST}), the Very Large Telescope (VLT) and \textit{Spitzer} as part of the CANDELS survey \citep{grogin+2011,koekemoer+2011}. In this work, we use the CANDELS GOODS-S multi-wavelength photometric catalogue \citep{guo+2013} that includes 17 bands: U Blanco/CTIO, U VLT/VIMOS, F435W, F606W, F775W, F814W and F850LP HST/ACS, F098M, F105W, F125W and F160W HST/WFC3, K$_{s}$ VLT/ISAAC, K$_{s}$ VLT/HAWK-I \citep{Fontana2014} and 3.6, 4.5, 5.8 and 8.0 $\mu$m Spitzer/IRAC \citep{Ashby2015}. The sources of this catalogue are detected in the F160W band and we correct all fluxes for galactic extinction. \par

The HAWK-I K band data available in the \cite{guo+2013} catalogue do not cover the full GOODS-S field, therefore we complement these observations with other K VLT/HAWK-I observations (P.I. N.Padilla) that are shallower than the observations in the \cite{guo+2013} catalogue in the same band but cover the full field. To extract photometry in this band, we run SExtractor \citep{bertin1996} using a 1.5$\sigma$ threshold over four contiguous pixels and an aperture of 3$^{\prime\prime}$. A low detection threshold is used to find a counterpart for all secure detections at 1.6$\mu$m. We correct the 3$^{\prime\prime}$ fluxes to total fluxes and ensure no offset with respect to \cite{guo+2013}. \par

The GOODS-S field has been observed also at mid- and far-IR wavelengths. In particular, we include in our data set observations taken with the Multiband Imaging Photometer  \citep[MIPS;][]{Rieke2004} on \textit{Spitzer} at $24 \, \rm \mu$m and observations done by the Photodetector Array Camera and Spectrometer \citep[PACS;][]{Poglitsch2010} on \emph{Herschel} \citep{Pilbratt2010} at 70$\mu$m, 100$\mu$m and 160$\mu$m as part of the PACS \citep{Poglitsch2010} \textit{Evolutionary Probe} (PEP, \citealt{Lutz2011}) and GOODS-\emph{Herschel} \citep{Rieke2011} programs. \par

\subsection{Sample selection and counterpart identifications}
Our main catalogue consists of 34930 galaxies that are part of the CANDELS catalogue and are detected in the F160W band, with the addition of HAWK-I K band observations that cover the full GOODS-S sample and observations in the mid- and far-IR with \textit{Spitzer} and \textit{Herschel}. The CANDELS catalogue can be considered to correspond to a mass-limited sample, because galaxies are selected in F160W and the mass-to-light ratio is quite stable at these wavelengths. On the other hand, the subsample with mid- and far-IR observations is biased towards highly star-forming and dusty galaxies.\par

First, we add the HAWK-I/K band observations to the CANDELS catalogue by cross-matching the catalogue with the wide HAWK-I/K band observations with a 1$^{\prime\prime}$ search radius. When more than one counterpart is present in the K-band ($\sim7\%$ of the catalogue), mainly because of the low threshold used for the source extraction in this band, we consider the brightest K band source as a counterpart, that is usually also the closest one.\par

For the \textit{Spitzer}/MIPS $24 \, \rm \mu$m catalogue we consider only sources brighter than 80 $\mu$Jy, for which the catalogue completeness is $\sim80\%$ and the spurious sources are less than 10$\%$ \citep[e.g.][]{Papovich2004}. In order to identify the optical counterpart of each $24 \, \rm \mu$m-detected source, we cross-match the MIPS catalogue and the CANDELS+HAWK-I/K catalogue using a 2$^{\prime\prime}$ searching radius respect to the HAWK-I/K source positions. Among the initial $24 \, \rm \mu$m-detected galaxies, 733 (98$\%$) have an optical-IR counterpart. The remaining 8 objects, detected at $24 \, \rm \mu$m, do not have any CANDELS counterpart within 2$^{\prime\prime}$. However, visual analysis of these sources shows that they could be associated with extended sources or multiple objects around the MIPS detection. We do not consider $24 \, \rm \mu$m fluxes when multiple optical counterparts are associated with the same $24 \, \rm \mu$m source. To conclude, 644 sources have photometry also at $24 \, \rm \mu$m, corresponding to the 88$\%$ of the total MIPS catalogue and $\sim2\%$ of the total CANDELS catalogue.\par

At far-IR wavelengths,  \citealt{Magnelli2013} derived two different catalogues from PEP observations: in one of them, sources are extracted blindly in the three bands and then they are crossmatched with a $24 \, \rm \mu$m-detected source catalogue; in the other one,  sources are extracted using the $24 \, \rm \mu$m source positions as prior. We cross-match our catalogue with both PEP catalogues with a radius of 1$^{\prime\prime}$ by comparing the $24 \, \rm \mu$m positions. Around 90$\%$ (585) of objects detected at $24 \, \rm \mu$m are also detected in at least one of the three PEP bands, in particular $\sim50\%$ of the $24 \, \rm \mu$m sources are detected at 70 $\mu$m, 86$\%$ at 100 $\mu$m and 78$\%$ at 160 $\mu$m.  \par

\section{Photometric redshifts and stellar masses}\label{sec:zphot_mass}

Our catalogue is different, in the optical and near-IR, from the \cite{guo+2013} catalogue for the presence of the HAWKI-K band observations which covers the full field. For this reason we derive both the photometric redshift and stellar mass using our own catalogue. In particular, to determine simultaneously the photometric redshift and the stellar mass of each galaxy, we run the code \textit{LePhare} \citep{arnouts+1999,ilbert+2006} to obtain the best SED templates based on a $\chi^{2}$ fitting procedure. We use \citet{BC03} templates considering solar metallicity, a range of  exponentially declining star formation histories and delayed star formation histories, both with different characteristic time scales $\tau$ from 0.1 to 10 Gyr, and ages ranging from 0.05 to 13.5 Gyr. We allow for redshift between 0 and 6, we apply the Calzetti et al. reddening law \citep{calzetti+2000} for the galactic extinction with $A_{V}$ values from 0 to 4  and we allow \textit{LePhare} to include nebular emission lines. We fit models by using all the optical bands and the two shortest-wavelength IRAC bands, because the two longest wavelengths have a lower S/N than the shortest wavelength bands and they could be contaminated by polycyclic aromatic hydrocarbon (PAH) emissions at low redshift. \par

\begin{figure}[!htbp]
\centering
\includegraphics[width=1\linewidth, keepaspectratio]{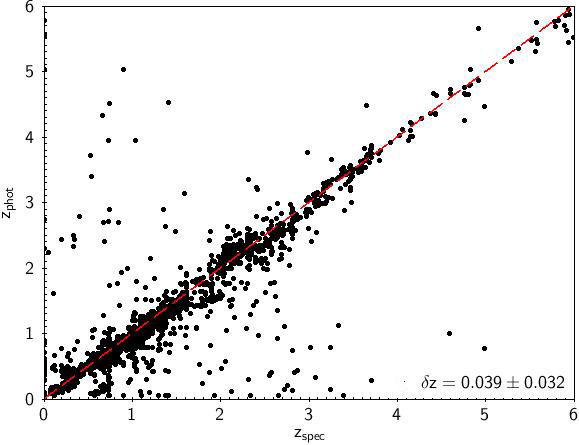}
\caption{Comparison between photometric redshifts of this work and spectroscopic redshifts presented in the literature \citep{Cristiani2000,Croom2001,Bunker2003,Dickinson2004,Stanway2004,Stanway2004b,Strolger2004,Szokoly2004,vanderWel2004,Doherty2005,Lefevre2005,Mignoli2005,Ravikumar2007,Popesso2009,Balestra2010,Silverman2010,Kurk2013,Vanzella2008,Morris2015}. The deviation is $\delta z=|z_{phot}-z_{spec}|/(1+z_{spec})=0.039\pm0.032$, excluding the 11$\%$  of outliers with $\delta z>0.15$.}
\label{fig:zspec_zphot}
 \end{figure}

We compare our photometric redshifts with spectroscopic redshifts available in the literature in the GOODS-S field. We use secure spectroscopic redshifts (\textit{z$_{spec}$}) present in the ESO public compilation in the Chandra Deep Field South \citep[CDFS;][]{Cristiani2000,Croom2001,Bunker2003,Dickinson2004,Stanway2004,Stanway2004b,Strolger2004,Szokoly2004,vanderWel2004,Doherty2005,Lefevre2005,Mignoli2005,Ravikumar2007,Popesso2009,Balestra2010,Silverman2010,Kurk2013,Vanzella2008} and spectroscopic redshifts from \citealt{Morris2015}, for a total of 2635 matches in the CANDELS catalogue.  The deviation of our photometric redshifts from the spectroscopic ones is $\delta z=|z_{phot}-z_{spec}|/(1+z_{spec})=0.039\pm0.032$ for the full sample, excluding outliers with $\delta z>0.15$ that correspond to $\sim11\%$ of the sample (Fig. \ref{fig:zspec_zphot}). These outliers are distributed in a wide range of stellar masses and SFRs and do not occupy specific regions of the SFR-M$^{*}$ plane, therefore they do not have significant influence on the results of this paper (Appendix \ref{sec:outliers}).\par

We divide our sample into three redshift bins, in order to study the position of galaxies on the SFR-M$^{*}$ plane:  0.5$\leqslant z<$1, 1$\leqslant z<$2 and 2$\leqslant z <$3. In these three redshift bins there are 6881, 9578 and 8004, respectively, making a total of 24463 galaxies. \par
 
In order to calculate the stellar mass completeness of our sample, we follow the empirical approach presented in \citet{Schreiber2015}. In particular, the CANDELS sample is selected at 1.6 $\mu$m, where the light-to-mass ratio is quite stable. Indeed, a tight relation is present between the stellar mass and the observed luminosity at 1.6 $\mu$m that can be described as a power-law plus a Gaussian scatter in the log space. We derive this relation and the related scatter directly from the data in different redshift bins. From the 3$\sigma$  detection limit in the F160W band it is possible to derive the corresponding 3$\sigma$ limit at each redshifts for the L$_{1.6\mu m/(1+z)}$. Then, we randomly extract points inside the Gaussian scatter of this relation in a wide range of stellar masses and luminosities and we derive the percentages of galaxies above the 3$\sigma$ detection limit in each 0.1 dex stellar mass bin. The 90$\%$ completeness limits are at log$_{10}(M^{*}/M_{o})$=7.35, 7.95 and 8.25 at 0.5$<z\leqslant$1, 1$<z\leqslant$2 and 2$<z\leqslant$3, respectively.

\section{Star formation rates}\label{sec:SFR}
For the full CANDELS sample, we do not use the SFR derived from the SED fitting, which highly depend on the assumed star formation history \citep{Pforr2012}, but we derive the SFR from the rest-frame UV continuum flux corrected for dust extinction using the \citet{calzetti+2000} reddening law and the A$_{v}$ value derived from the SED fitting. Depending on the redshift, we use fluxes in the U-band, F435W, F606W or F775W band. These UV fluxes are then converted to SFR using the relation derived by \cite{Kennicutt1998}:
\begin{equation}\label{Kennicutt_UV}
SFR [M_{\odot} yr^{-1}] = 1.4 \times 10^{-28} L_{\nu} [ergs~s^{-1} Hz^{-1}]
.\end{equation}\par

For the subsample of galaxies detected at $24 \, \rm \mu$m, we also derive the SFR from the dust-processed stellar light in the IR with the addition of the UV-based SFR explained previously, but not corrected for dust extinction. In particular, the total IR luminosity from 3 to 1000 $\mu$m is related to the SFR, as it was found by \cite{Kennicutt1998}:
\begin{equation}\label{Kennicutt}
SFR[M_\odot yr^{-1}]=\frac{L_{FIR} [L_\odot]}{5.8 \times 10^9}
.\end{equation}
However, we do not observe directly the total IR luminosity and several methods are present in the literature to estimate this total luminosity from a monochromatic luminosity. In this work, we consider five different methods to derive the total IR luminosity: 
\begin{itemize}
  \item the relation by \cite{Chary2001} between IR and monochromatic luminosities, using the observed $24 \, \rm \mu$m flux;
  \item the relation by \cite{Bavouzet2008} between IR and monochromatic luminosities, using the observed flux at $24 \, \rm \mu$m;
  \item the relation by \cite{Bavouzet2008} between IR and monochromatic luminosities, using the observed flux at 70 $\mu$m;
  \item the relation by \cite{Rieke2009} between IR and observed flux at $24 \, \rm \mu$m;
  \item a direct fit to the available IR observations, from $24 \, \rm \mu$m to 160 $\mu$m, by using IR templates presented in \cite{Rieke2009}.
\end{itemize}
When IR observations are not available, we consider SFR derived from the UV and corrected for dust-extinction. \par

\subsection{K-corrections}
Every method explained in the previous paragraph converts a monochromatic luminosity at a reference wavelength into total IR luminosity. Because the IR data are available in specific bands, we need to apply a k-correction to obtain the luminosity at the reference wavelength of each method.
We start from models of pure starburst galaxies \citep[][]{Rieke2009} that have IR luminosities from 10$^{9.75}$ to 10$^{13}$ L$_{\odot}$. We convolve each model with the $24 \, \rm \mu$m, 70$\mu$m, 100$\mu$m and 170$\mu$m filters. We then calculate the k-correction by dividing the observed flux by the flux at the reference wavelength. We use this ratio to calculate the total IR luminosity and we compare it with the one associated to the used template. We iterate the process until the output IR luminosity corresponds to the IR luminosity of the used model.

\subsection{IR luminosity derivation}\label{sec:LIR}
We derive total IR luminosities by using four different relations between monochromatic and total IR luminosities and, independently, by fitting SED models to the total fluxes at $24 \, \rm \mu$m, 70 $\mu$m, 100 $\mu$m and 160 $\mu$m. \par
\begin{enumerate}
\item \citet[][here after C$\&$E]{Chary2001} studied different samples of nearby galaxies and for galaxies with L$_{IR}>10^{10}$ L$_{\odot}$ they derived three relations between total IR luminosity and monochromatic luminosity at 6.7 $\mu$m, 12 $\mu$m and 15$\mu$m:
\begin{equation} \label{C&E}
L_{IR} = 
  \begin{cases} 
    11.1^{+5.5}_{-3.7} \times L^{0.998}_{15 \mu m} & : z\leqslant 0.8\\
    0.89^{+0.38}_{-0.27} \times L^{1.094}_{12 \mu m} & : 0.8>z \leqslant 1.6 \\
    4.37^{+2.35}_{-2.13} 10^{-6} \times L^{1.62}_{6.7 \mu m} & : z>1.6.   
  \end{cases}
\end{equation}
Close to each relation we show the redshift range in which we implement it, in order to apply the smallest k-correction possible from the flux at $24\mu$m and the flux at the wavelength considered in each relation.\par
\item In \citet{Bavouzet2008}, the total IR luminosity is related to the monochromatic luminosity at 8, 24 and 70 $\mu$m by:
\begin{equation} \label{Ba}
L_{IR} = 
  \begin{cases} 
    377.9 \times L^{0.83}_{8 \mu m}\\
    6856 \times L^{0.71}_{24 \mu m}\\
   7.90 \times L^{0.94}_{70 \mu m.}    
  \end{cases}
\end{equation}
In this case we use observations at $24 \, \rm \mu$m (here after Ba24) and 70 $\mu$m (here after Ba70), when available. In order to apply the smallest k-correction possible, in Ba24, we use the  first relation at z>0.5 and the second one for closer objects. Similarly, in Ba70, we use the third relation at z$\leqslant$0.5, the second one at 0.5$<$z$\leqslant$5 and the first one otherwise. \par
\item In \citet{Rieke2009} (here after Ri09), SFR is directly related to $24 \, \rm \mu$m observed flux by:
\begin{equation}
log(SFR)= A(z) + B(z)(log(4 \pi D_L^2f_{24,obs})-53)
,\end{equation}
where D$_{L}$ is the luminosity distance and values for A and B are listed in the same paper for different redshifts below 3. We do not apply this relation at z$>$3.  We derive IR luminosity for Ri09 by reversing Eq. \ref{Kennicutt}.\par
\end{enumerate}
Finally, when at least two observations are present in the IR, we use again the templates from \citet{Rieke2009} to directly fit observations at 24, 70, 100 and 160 $\mu$m using a $\chi^{2}$ fitting procedure to estimate the best solutions among all templates. However, among different SED templates presented in the literature there is a large scatter on the PAH features strength. In order for our results to not be too dependent on the SED template
used, we multiply by two the error bars associated to $24 \, \rm \mu$m fluxes, which contained PAH features between z$\sim$0.8 and z$\sim$2.8. We also derive L$_{IR}$ by SED fitting without including observations at $24 \, \rm \mu$m, but the difference is less than 20$\%$ for 99$\%$ of the sample with observations in at least two bands. \par

In order to estimate the errors associated with the IR luminosity for each monochromatic relation, we consider both the propagation of the flux and photometric redshift errors inside each relation and the intrinsic scatter of each relation. The mean errors are 0.35 dex for C$\&$E, 0.32 dex for Ba24, 0.43 dex for Ba70 and 0.21 dex for Ri09. The mean error of Ba70 is higher than the mean error of Ba24, because S/N associated with the flux at 70$\mu$m are generally smaller than the ones at $24 \, \rm \mu$m.  \par
On the other hand, in order to estimate IR luminosity errors for the SED fitting technique, we analyse the probability distribution, P$(L_{IR})\propto e^{-\chi^2/2}$, and we derive the full width half maximum associate to this distribution for each galaxy. The mean IR luminosity error associated with the SED fitting method is 0.09 dex. \par
\begin{figure*}[]
\centering
 \includegraphics[width=1\linewidth, keepaspectratio]{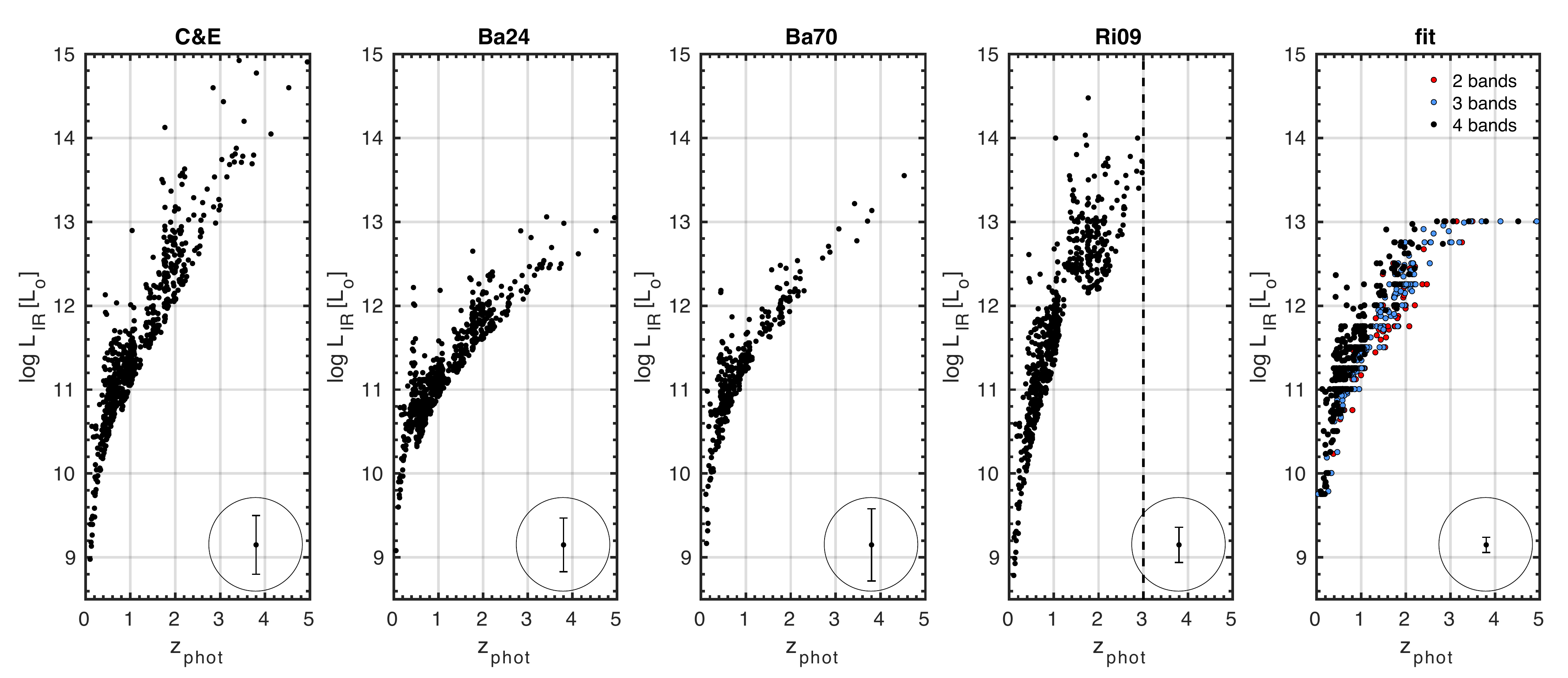}
 \caption{Photometric redshift compared to the IR luminosity derived using different methods: \textit{From left to right:} C$\&$E, Ba24, Ba70, Ri09 and IR SED fitting. In the fourth panel, the dashed vertical line corresponds to the redshift limit below which the relation to derive IR luminosity is calibrated. In the third panel, open symbols are upper limits. In the fifth panel, points are colour-coded depending on the number of bands with detection: (\textit{black}) 4 bands, (\textit{blue}) 3 bands and (\textit{red}) 2 bands. On the bottom part of each plot, it is shown the mean error bar associate to each method.}
 \label{fig:LIR}
 \end{figure*}

In Figure \ref{fig:LIR}, we show the relation between IR luminosity and photometric redshift for each conversion formulae used to derive the bolometric infrared luminosity L$_{IR}$. Both Ri09 and C$\&$E methods produce $L_{IR}>10^{14} L_{\odot}$, that are not present in other methods. The IR luminosities derived with Ba24 and Ba70 have the shallowest increment with redshift among the methods tested here and, in particular, IR luminosities remain below $10^{13} L_{\odot}$ for the majority of cases. IR luminosities derived fitting the IR photometry are always L$<10^{13} L_{\odot}$, because this is the maximum luminosity present in the used templates. Differences among the derived IR luminosities are due to small variations in the K-corrections, but mainly to differences among the used conversion formulae because of the different galaxy samples used to derive them. \par
 
\begin{figure}[]
\centering
 \includegraphics[width=1\linewidth, keepaspectratio]{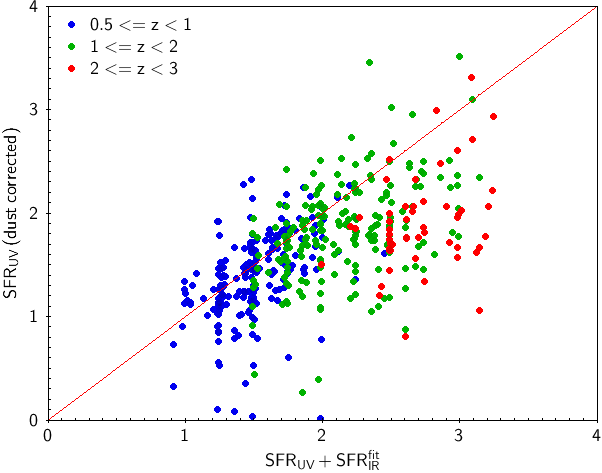}
 \caption{SFR derived from the fit of the IR SED and the uncorrected UV SFR (SFR$_{IR}^{fit}+$SFR$_{UV}$) compared to dust-corrected UV SFR, for galaxies which have both quantities. Galaxies are divided in the three redshift bins analysed in this work: 0.5$\leqslant z<$1 (\textit{blue}), 1$\leqslant z<$2 (\textit{green}) and 2$\leqslant z <$3 (\textit{red}). }
 \label{fig:SFRUV_IR}
 \end{figure}

 In Figure \ref{fig:SFRUV_IR}, we show the comparison between UV, dust-corrected SFR and SFR$_{IR}+$SFR$_{UV}$, where the IR SFR is derived by fitting the IR SED the UV SFR is not corrected for dust extinction, for the galaxies which have mid- and far-IR observations. The two SFRs are, overall, consistent, in particular at 0.5$\leqslant z<$1. However, the SFR$_{UV}$ tends to be underestimated at high SFR and redshifts, particularly at 2$\leqslant z <$3. This comparison is different when comparing the SFR$_{UV}$ with SFR$_{IR}$ derived with other methods, but these quantities remain in overall agreement. \par

\section{Analysis of sources on the $SFR-M^{*}$ plane}\label{sec:MSfit}
We divide our sample into three redshift bins, 0.5$\leqslant z<$1, 1$\leqslant z<$2 and 2$\leqslant z <$3, in order to study the SFR-M$^{*}$ plane and the MS slope at different redshifts, and we use two different fitting techniques to derive it.\par

Firstly, we remove quenched galaxies from our sample, selecting only galaxies with log$_{10}(sSFR/yr^{-1})>$-9.8 and we derive the MS by fitting a linear regression using all non-quenched galaxies in each redshift bin. This method is similar to what is usually found in the literature and it allows us to make a direct comparison with previous work. \par

Secondly, we follow \citet{Renzini2015} who analysed the 3D SFR-M$^{*}$ relation (SFR vs. stellar mass vs. number of galaxies inside each SFR-M$^{*}$ bin) and identified two peaks, one for quenched galaxies and one for MS galaxies, and defined the MS as the ridge line of the MS distribution. This definition has the advantage of not requiring a pre-selection to derive the MS position and, at the same time, is not influenced by the presence of quenched galaxies when deriving the MS. Following their approach, we divide galaxies into stellar mass bins of 0.25 dex and we fitted the sSFR distribution inside each stellar mass bin with three different Gaussian distributions, corresponding to the quenched galaxies (QG), the MS galaxies, and the starburst galaxies (SB). In \citet{Renzini2015} the third component (starburst galaxies) was not evident because this population is rare at the low redshifts that they considered (0.02 $< z <$ 0.085).  We use the sSFR instead of the SFR, because the three components are more clearly separated in the sSFR distribution than in the SFR one. Other studies have analysed the sSFR distribution at different stellar masses \citep[e.g.][]{Ilbert2015}, but they did not model this distribution to separate among SF, quenched, and starburst galaxies, and derive the MS.
Finally, we fit an orthogonal distance regression (ODR) to the derived peak positions of the three Gaussian components in each stellar mass bin, in order to take into account the stellar mass bin size as well as the errors associated with each Gaussian peak position. The sequences derived by using this method are sensitive to the adopted stellar mass bin, however, the results are all consistent. \par

\subsection{Single fitting to $SFR-M^{*}$ plane}\label{Linearfit}
\begin{figure*}[]
\centering
 \includegraphics[width=1\linewidth, keepaspectratio]{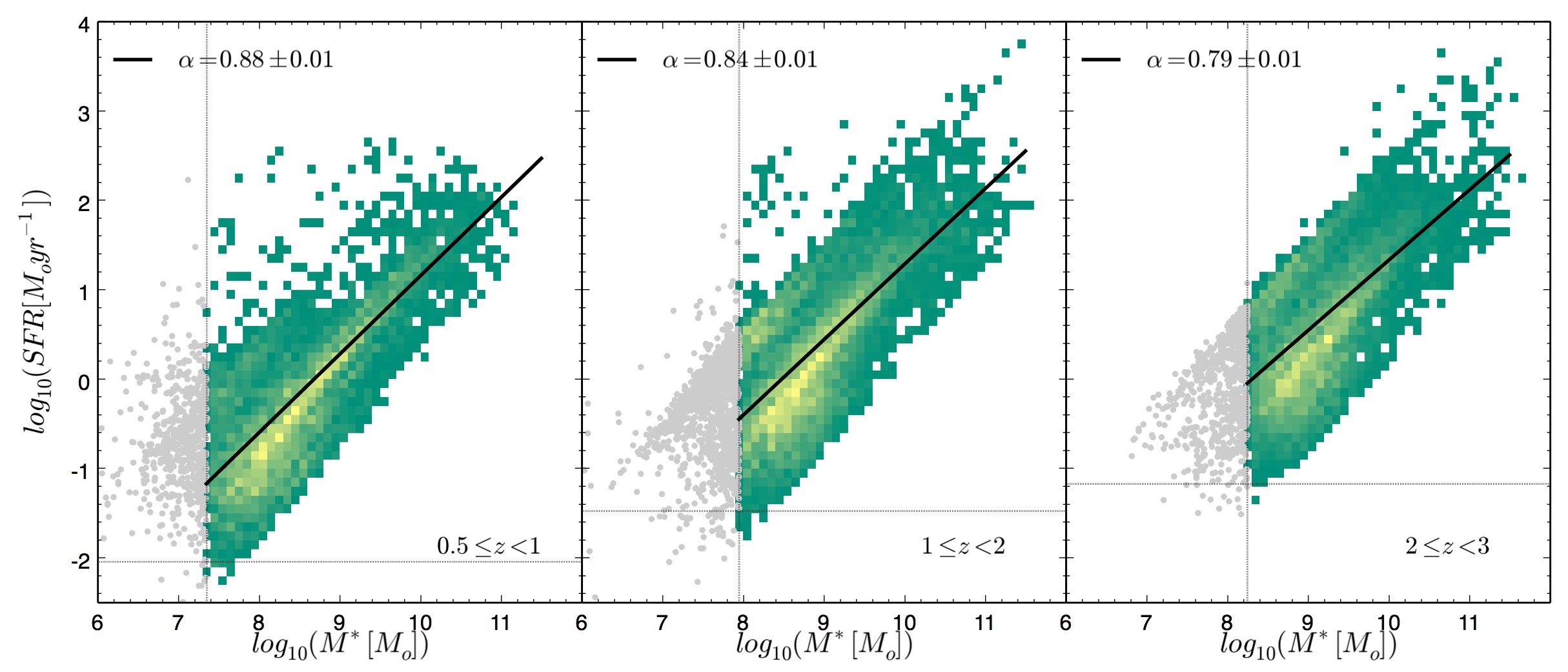}
 \caption{SFR-M$^{*}$ plane with the best linear fits to all SF galaxies at different redshifts: 0.5$\leqslant z <$1 (\textit{left}), 1$\leqslant z <$2  (\textit{center}) and 2$\leqslant z <$3  (\textit{right}). The slope of each sequence is reported on the top-left of each panel. In the background there are all the SF galaxies in our sample above the stellar mass completeness, divided into SFR-M$^{*}$ bins of 0.1$\times$0.1 dex. Each bin is colour-coded depending on the number of galaxies inside each bin, from green to yellow in a linear scale. The vertical black dotted lines are the 90$\%$ completeness limit in stellar mass in each redshift bin; galaxies below this limit are plotted as grey points and are not included in the fit. The horizontal dotted black lines are the 3$\sigma$ detection limit in SFR in each redshift bin.}
 \label{fig:MS_linfit}
 \end{figure*}

Here we discuss the SFR-M$^{*}$ plane derived using a single linear regression fitting to all SF galaxies, as is commonly found in the literature.
In Figure \ref{fig:MS_linfit} we show the linear regression for all SF galaxies, which are selected by log$_{10}(sSFR/yr{-1})>$-9.8.  The slopes that we derive are 0.88$\pm$0.01, 0.84$\pm$0.01 and 0.79$\pm$0.01, at 0.5$\leqslant z <$1, 1$\leqslant z <$2 and 2$\leqslant z <$3, respectively. \par

\begin{figure*}[]
\centering
 \includegraphics[width=1\linewidth, keepaspectratio]{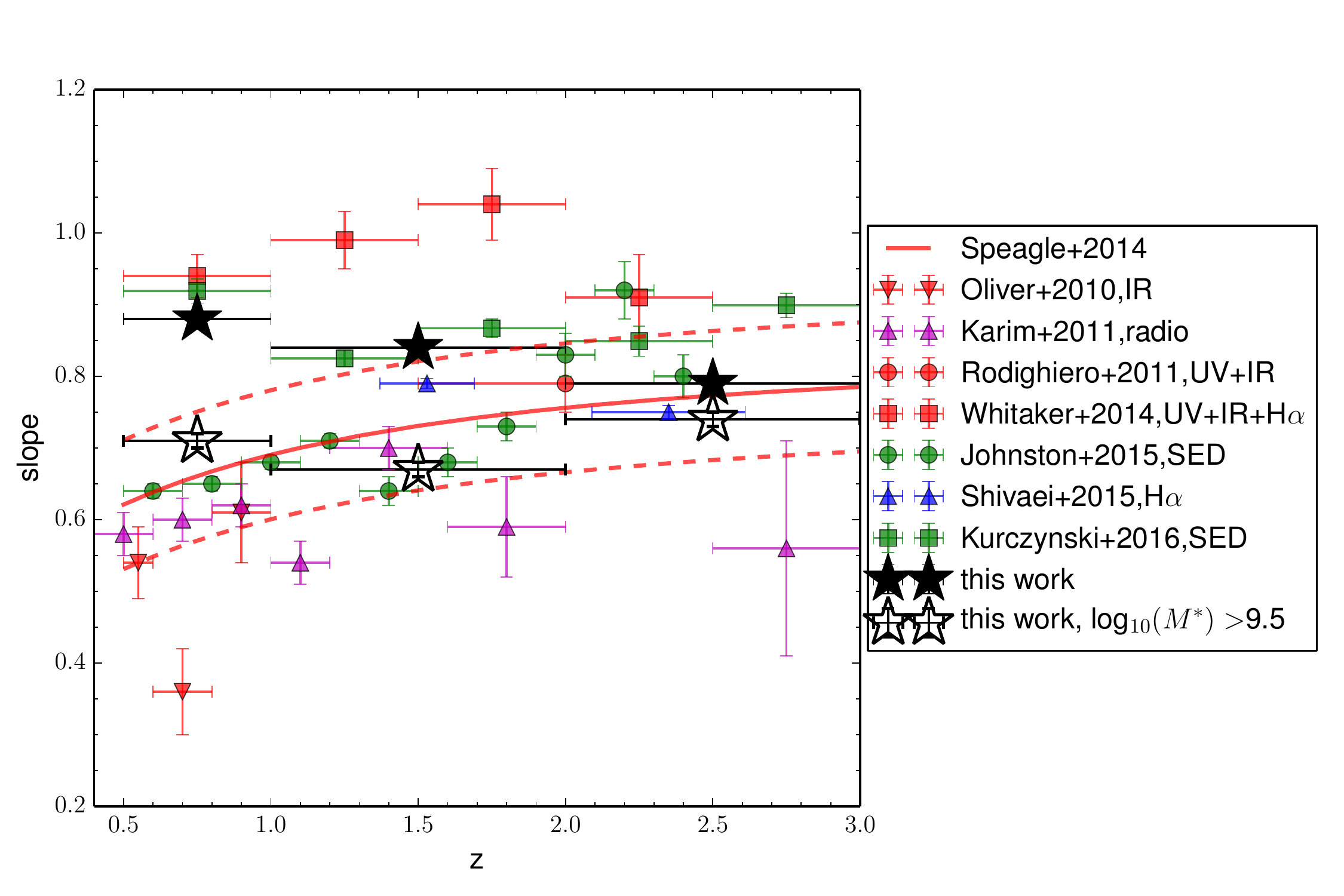}
 \caption{Comparison between some MS slopes present in the literature and the slopes derived in this work using all SF galaxies (\textit{black stars}) and only SF galaxies with log$_{10}(M^{*}/M_{o})>$9.5 (\textit{empty stars}). All other slopes are taken from: \cite{Oliver2010} (\textit{red triangles}), \cite{Karim2011} (\textit{magenta triangles}), \cite{Rodighiero2011} (\textit{red circles}), \citep{Whitaker2014} (\textit{red squares}), \cite{Johnston2015} (\textit{green circles}), \cite{Shivaei2015} (\textit{blue triangles}) and \cite{Kurczynski2016} (\textit{green squares}). The red continuous line shows the relation derived by \cite{Speagle2014} and the dotted red lines are the median inter-publication scatter around the fit.}
 \label{fig:MS_CANDELS}
 \end{figure*}

Figure \ref{fig:MS_CANDELS} shows the comparison between our derived slopes and some values present in the literature. The scatter among different methods is very large and our values are within this range. In particular, our values are above the values derived by \cite{Speagle2014}, which combine results from numerous previous works, and are more similar to recent values derived by \cite{Kurczynski2016}. However, in \cite{Speagle2014} the MS is analysed for log$_{10}(M^{*}/M_{o})>$9.5, therefore we re-derive the MS in a similar stellar mass range to compare with our work. The derived slopes are 0.71$\pm$0.01, 0.67$\pm$0.01 and 0.74$\pm$0.01 at 0.5$\leqslant z <$1, 1$\leqslant z <$2 and 2$\leqslant z <$3, respectively, which are consistent with the median inter-publication scatter around the relation found by \cite{Speagle2014}. This analysis makes evident to which extent the derived MS slope is influenced by the considered stellar mass range. \par

Using IR-based SFR for the MS, we find that the slopes remain almost unchanged. This is because the galaxies at low-intermediate stellar masses are dominating the fits with this method because they are the most numerous, and for these galaxies the SFR is derived from the UV, because there are no observations in the IR. \par
 
\subsection{Multi-Gaussian decomposition of the SFR-M$^{*}$ plane}
Here we discuss the SFR-M$^{*}$ plane by separating SB, SF and QG galaxies fitting the sSFR distribution with three Gaussian components  in different stellar mass and redshift bins. \par

In Figure \ref{fig:Gauss0} we show the best Gaussian least squares fits of the sSFR distribution for our sample, in the three considered redshift bins and for stellar masses between log$_{10}(M^{*}/M_{o})=$7.5 and 11.25. We analyse only stellar mass bins with more than 50 galaxies. The three Gaussian components used in the fit, one for each galaxy mode (QG, MS and SB), have different relative positions and relative heights, changing both with stellar masses and redshift. In particular, at high stellar masses the three components are more separated one from the other than at low stellar masses. \par
The SB component in the fit is necessary to describe the observed sSFR distribution in some stellar mass and redshift ranges where SB are not negligible, as analysed in Appendix \ref{sec:AIC} by comparing the Akaike Information Criterion \citep[AIC;][]{Akaike1973} for a fit with and without the SB component.

\subsubsection{The three modes in the SFR-M$^{*}$ plane}
\begin{figure*}[]
\centering
 \includegraphics[width=1\linewidth, keepaspectratio]{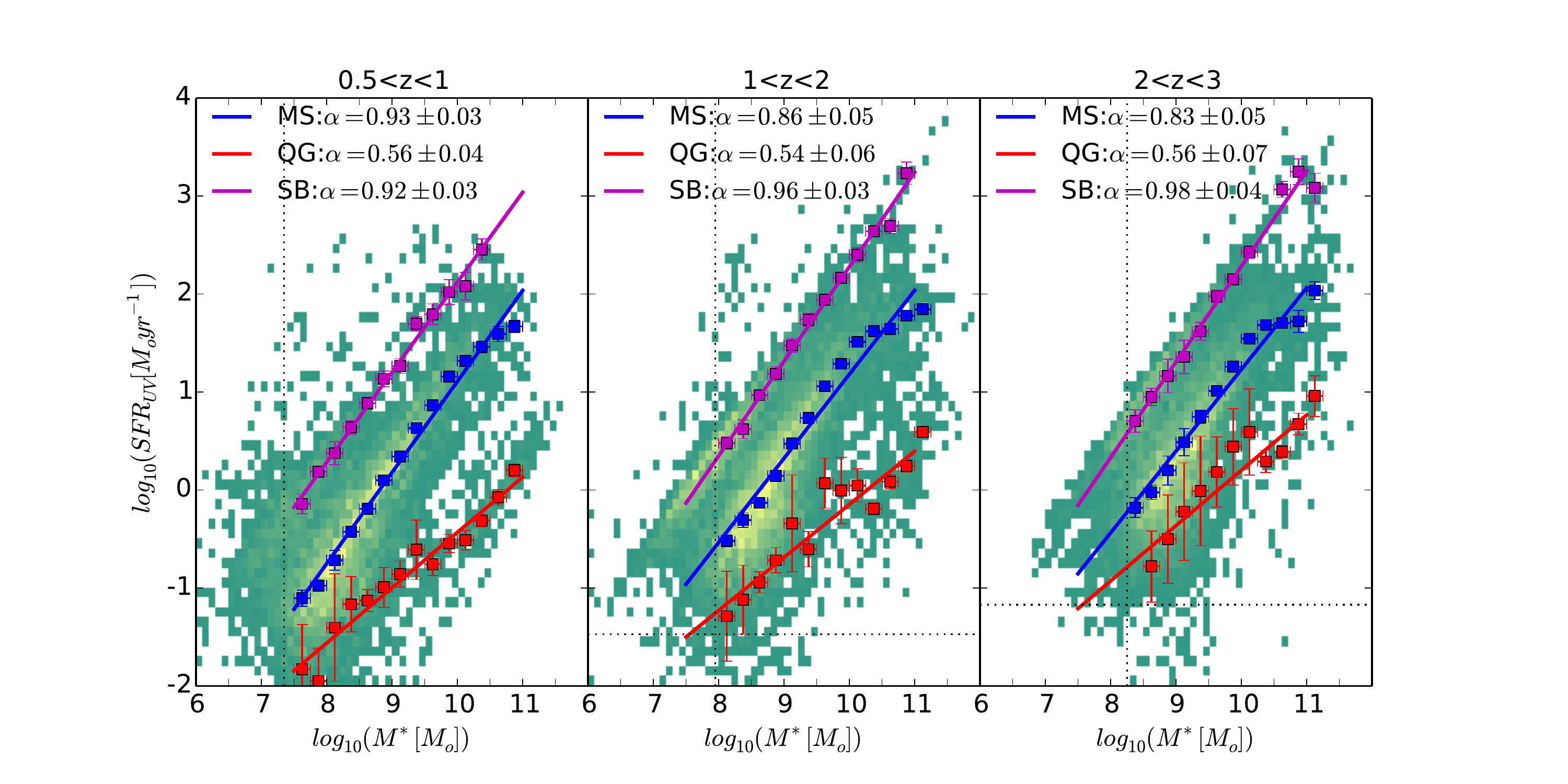}
 \caption{SFR-M$^{*}$ plane with the best ODR fits for the three galaxy modes: quenched galaxies (\textit{red line}), MS galaxies (\textit{blue line}) and SB galaxies (\textit{magenta line}). The slope of each sequence is reported in the top-left of each panel. Squared symbols are the Gaussian peak positions of the three components in each stellar mass bin with the associated error in the position. In the background there are all galaxies in our sample divided into SFR-M$^{*}$ bins of 0.1$\times$0.1 dex. Each bin is colour-coded depending on the number of galaxies inside each bin, from green to yellow in a linear scale. The vertical black dotted lines are the 90$\%$ completeness limit in stellar mass in each redshift bin. The horizontal dotted black lines are the 3$\sigma$ detection limit in SFR in each redshift bin.}
\label{fig:MS_CANDELS_SB}
\end{figure*}

For each galaxy mode, we derive the Gaussian parameters from the non-linear least squares fit. We calculate the errors associated to each Gaussian parameter by adding in quadrature the errors estimated from the fit and the errors derived by performing a bootstrap analysis. Then, we use the peak positions and its error to fit an ODR and derive the sequence for SB, SF and quenched galaxies, as shown in Figure \ref{fig:MS_CANDELS_SB}. The SB sequence has a slope close to unity up to log$_{10}(M^{*}/M_{o})=$11.  The MS has a slope between 0.93$\pm$0.03 and 0.83$\pm$0.05, decreasing with the redshift, but a flattening is present at high stellar masses, in particular at z$<$2. This flattening has been observed previously in other works with different methods \citep{Whitaker2014,Tomczak2016} and, because this flattening is present in our data after selecting only SF galaxies, it is not due to contamination by quenched galaxies. However, the inclusion of IR observations influences this flattening, as is explained in detail in Section \ref{sec:Gauss_IR}. When fitting only data points with log$_{10}(M^{*}/M_{o})<$10.25, below the high mass bending, the slope is close to unity at all redshifts and with a mild evolution. On the other hand, the QG sequence is less defined than the other two, with the high-mass part (log$_{10}(M^{*}/M_{o})>$10) well separated from the MS and the low-mass part very close to the MS, resulting in a slope between 0.54 and 0.56. A similar behaviour has been observed with low-z galaxies by \cite{Renzini2015}. \par 

\subsubsection{The evolution of the starburst and quenched fractions}
\begin{figure}[]
\centering
 \includegraphics[width=1\linewidth, keepaspectratio]{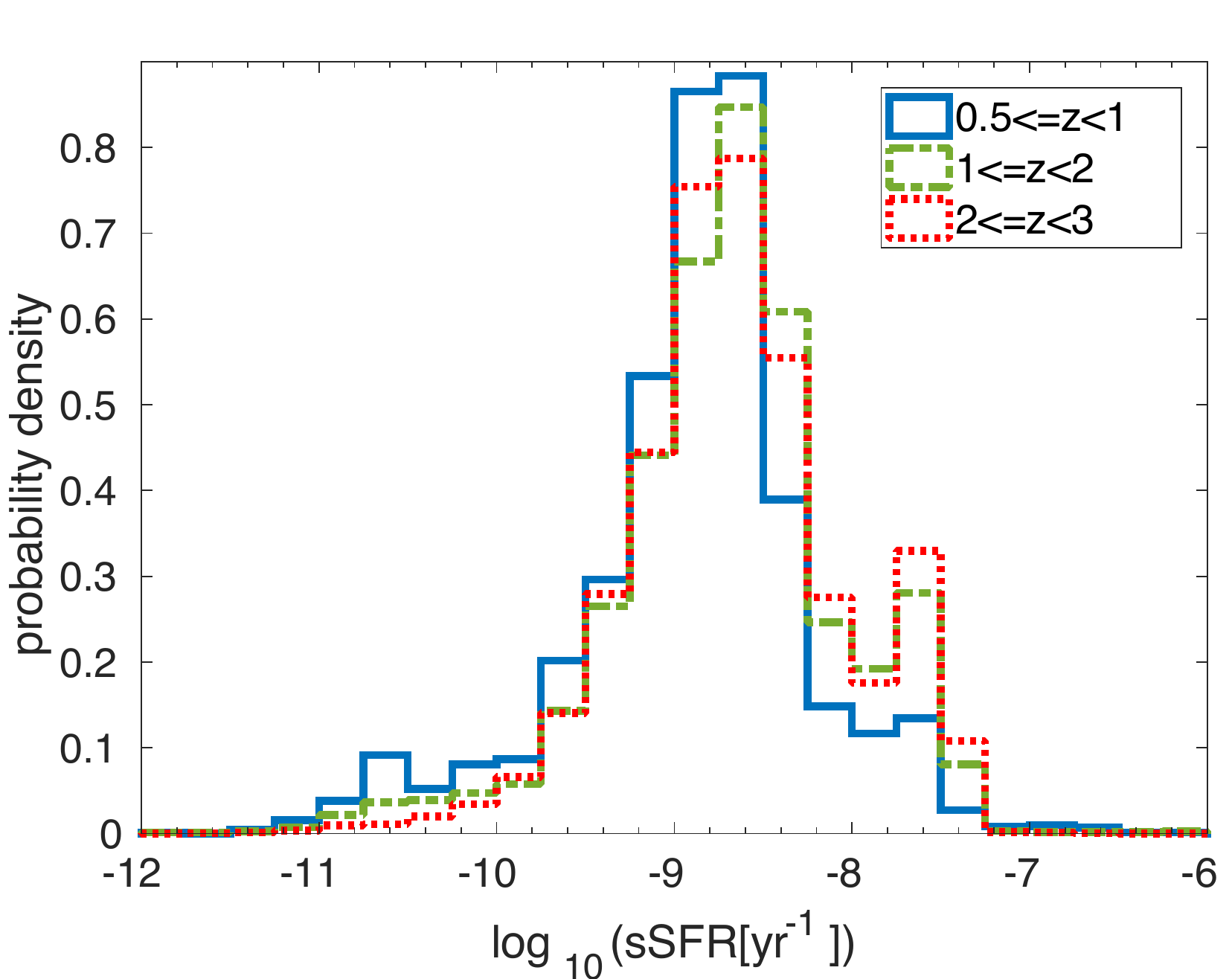}
 \caption{sSFR probability density distribution of all CANDELS galaxies in different z, only considering sources above stellar-mass completeness.}
\label{fig:sSFR_tot}
\end{figure}

In Figure \ref{fig:sSFR_tot} we show the sSFR probability density distribution in the three analysed redshift bins, above each stellar mass completeness. In this plot it is already evident that the number of starburst and quenched galaxies varies with redshift. This evolution could be partially due to selection effects for the QG, because they could be below the detection limit. On the other hand, there is a clear increment in the number of SB between z$\sim$0.75 and z$\sim$1.5. \par

\begin{figure}[]
\centering
 \includegraphics[width=1\linewidth, keepaspectratio]{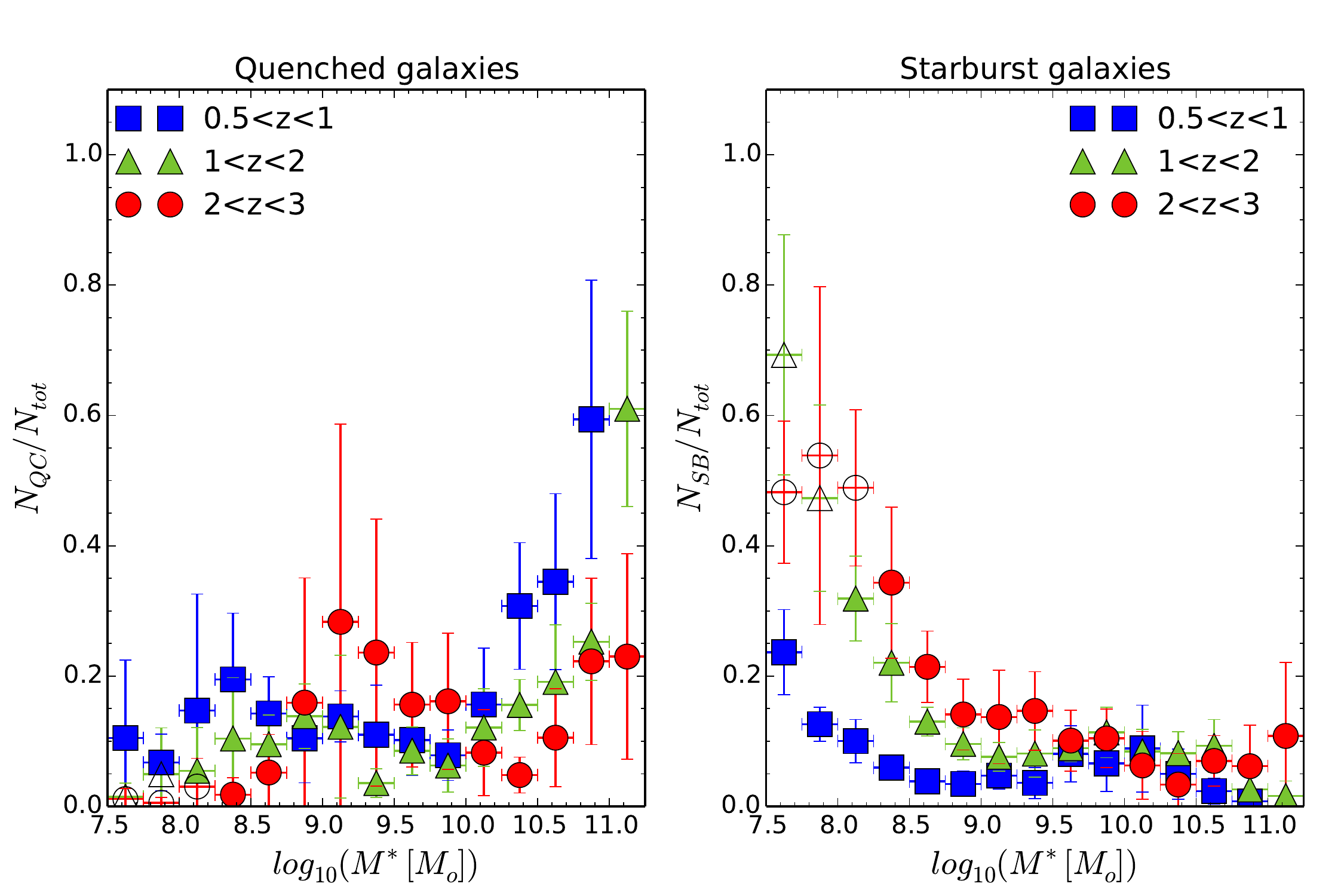}
 \caption{\textit{Left:} ratio between QG and the total number of galaxies in our sample at different stellar masses. \textit{Right:} ratio between SB and the total number of galaxies in our sample at different stellar masses. Points are divided into three different redshift bins: 0.5$\leqslant z<$1 (\textit{blue squares}), 1$\leqslant z<$2 (\textit{green triangles}) and 2$\leqslant z <$3 (\textit{red circles}). Empty symbols correspond to stellar-mass bins where the sample is below the 90$\%$ mass completeness. }
 \label{fig:Gratios}
 \end{figure}

To further investigate this evolution, we study the fraction of the three different galaxy modes at different stellar masses and redshifts. In particular, in Figure \ref{fig:Gratios} we show the ratio between the number of starburst (quenched) galaxies with respect to the total number galaxies for different stellar masses and redshift bins.  Galaxies in our sample are classified depending on the three Gaussian fits of the sSFR distribution in each stellar mass bin. The number of galaxies in each mode is derived by integrating the associated Gaussian component in each stellar mass and redshift bin. The errors associated to each fraction are derived propagating the Gaussian parameter errors, which are obtained adding in quadrature the errors derived from the fit and the ones derived from the bootstrap analysis. \par

In general, MS galaxies are the dominant mode, however, at log$_{10}(M^{*}/M_{o})>$10.5 and z$<$2, quenched galaxies are $>$20$\%$. At log$_{10}(M^{*}/M_{o})>$10, the number of QG decreases with increasing redshift, while it is almost constant at low-intermediate stellar masses ($\lesssim20\%$). Above this stellar mass (log$_{10}(M^{*}/M_{o})\sim$10), QG are also more separated from the MS, while below they are very close to the MS and are not a clear separate peak. This could be a stellar mass of transition between two different dominant quenching mechanisms. A dip is present in the fraction of QG at log$_{10}(M^{*}/M_{o})\sim$10, however, large error bars are presented at low stellar masses due to the proximity of QG and the MS. Therefore, it is not possible to understand if this dip is due to uncertainties or it is an underpopulated region between two different quenching modes. At z$>$1, the fraction of QG decreases at log$_{10}(M^{*}/M_{o})<$9, however, this fraction is underestimated in this redshift and stellar mass range because QG are below the 90$\%$ stellar mass completeness. \par

On the other hand, the fraction of SB galaxies increases with the redshift, particularly at log$_{10}(M^{*}/M_{o})<$9.  Moreover, this fraction increases by $\gtrsim$2 between redshift $\sim$0.75 and $\sim$1.5. The fraction of SB at the stellar mass range of log$_{10}(M^{*}/M_{o})>$10.5 could be contaminated by active galactic nuclei (AGN). However, these SB galaxies, when detected at $24 \, \rm \mu$m, do not show extreme IR fluxes and, moreover, only one galaxy in each stellar mass bin is classified as AGN at X-rays \citep{Luo2017}. Overall, looking at the total mass range above the 90$\%$ completeness (log$_{10}(M^{*}/M_{o})=$8.25-11.25), SB galaxies are $5\pm1\%$ of the total number of galaxies at 0.5$\leqslant z <$1, $12\pm1\%$ at 1$\leqslant z <$2, and $16\pm1\%$ at 2$\leqslant z <$3.  \par

For comparison with other works, \cite{Rodighiero2011} defined a starburst as a galaxy with sSFR 2.5$\sigma$ above the MS and found that they correspond the 2-3$\%$ of galaxies with log$_{10}(M^{*}/M_{o})=$10-11.5 at 1.5$\leqslant z <$2.5. In our work the SB fraction is generally higher in the same stellar mass range, probably because of the different definitions of starburst. Indeed, applying a cutoff at a defined sSFR does not take into account wings in the SB distribution as well as variation in the slopes or in the separation between the MS and the SB sequence with stellar masses. In agreement with our work, \cite{Sargent2012}, by using the same data of \cite{Rodighiero2011} but a technique similar to the one use in this work, found a higher fraction of SB galaxies than \cite{Rodighiero2011}.\par

\citet{Caputi2017} found that $\sim15\%$ of all galaxies with log$_{10}(M^{*}/M_{o})>$9.2 at z$=4-5$ are SB. Although at these higher redshifts, SB galaxies are characterised by much higher sSFR. This fraction is slightly higher than the fraction of SB galaxies we found at z$=$2-3 (11$\pm$1$\%$) in the same stellar-mass range, showing that the evolution of the fraction of SB in this stellar-mass range seems to continue up to z$=$5.

\begin{figure*}[!htbp]
\centering
\includegraphics[width=1\linewidth, keepaspectratio]{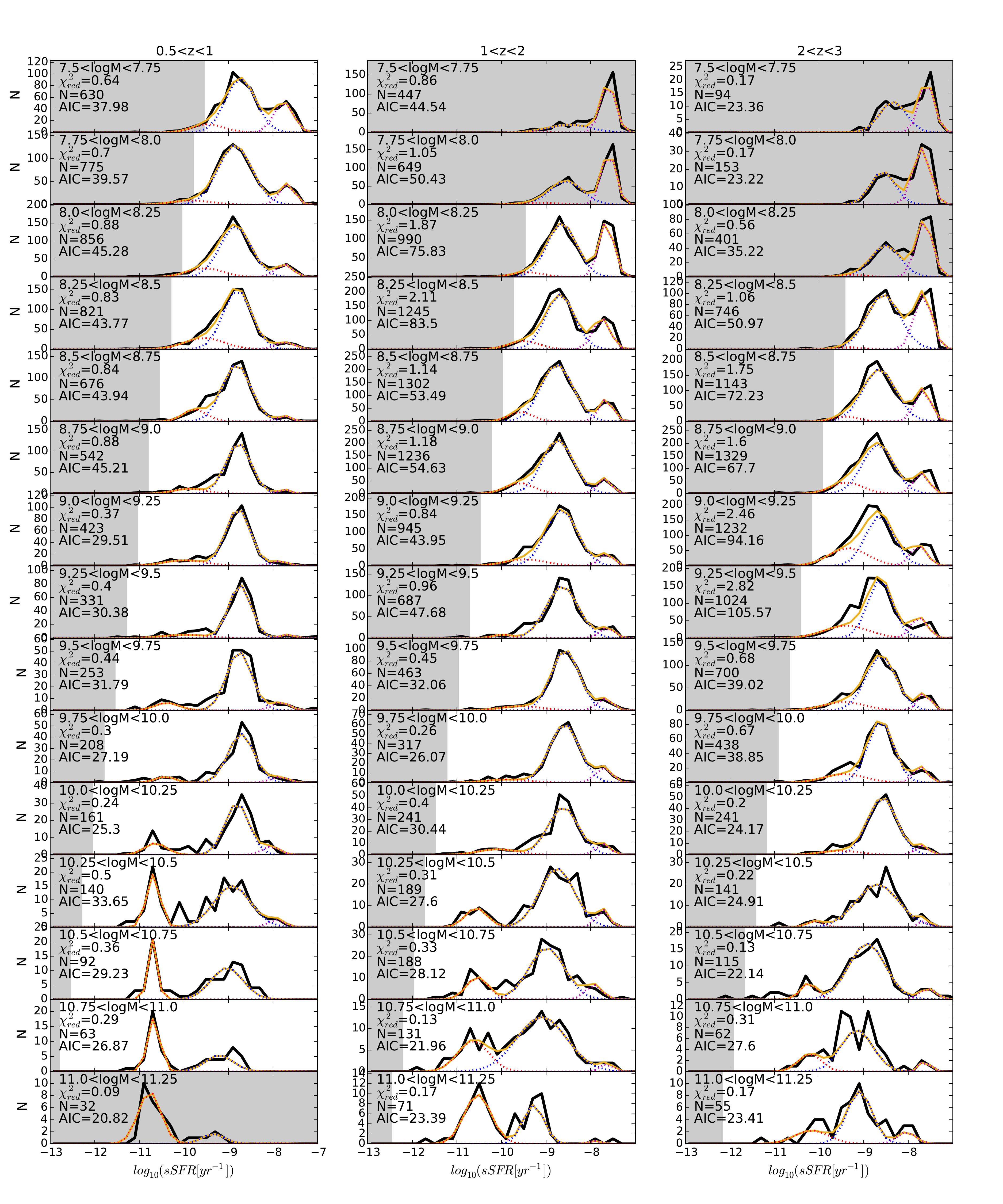}
\caption{sSFR distribution of the CANDELS sample in different z and stellar-mass bins. Each column shows a different redshift bin: 0.5$\leqslant z<$1 (\textit{left}), 1$\leqslant z<$2 (\textit{central}) and 2$\leqslant z <$3 (\textit{right}). Each row shows a different stellar-mass bin, with a width of 0.25 dex, from 7.5$ \leqslant log_{10}(M^{*})<$7.75 (\textit{top}) to 11$ \leqslant log_{10}(M^{*})<$11.25  (\textit{bottom}). Black thick continuous lines show the sSFR distribution and the yellow thin continuous lines are the best fitted models. Dotted lines are the different Gaussian components: quenched galaxies (\textit{red}), MS galaxies (\textit{blue}) and starburst galaxies (\textit{magenta}). The $\chi_{red}^{2}$ and AIC values of each fit and the number of galaxies in each bin are shown in the top left of each panel. Grey areas show the sSFRs and stellar masses for which this sample is not complete (below the 3$\sigma$ limit in sSFR or below the 90$\%$ completeness in stellar mass), or has statistics that are too low (N$<$50).}
\label{fig:Gauss0}
 \end{figure*}
 
 \subsubsection{Impact of using different IR conversion formulae to derive the bolometric IR luminosity}\label{sec:Gauss_IR}
\begin{figure*}[]
\centering
 \includegraphics[width=1\linewidth, keepaspectratio]{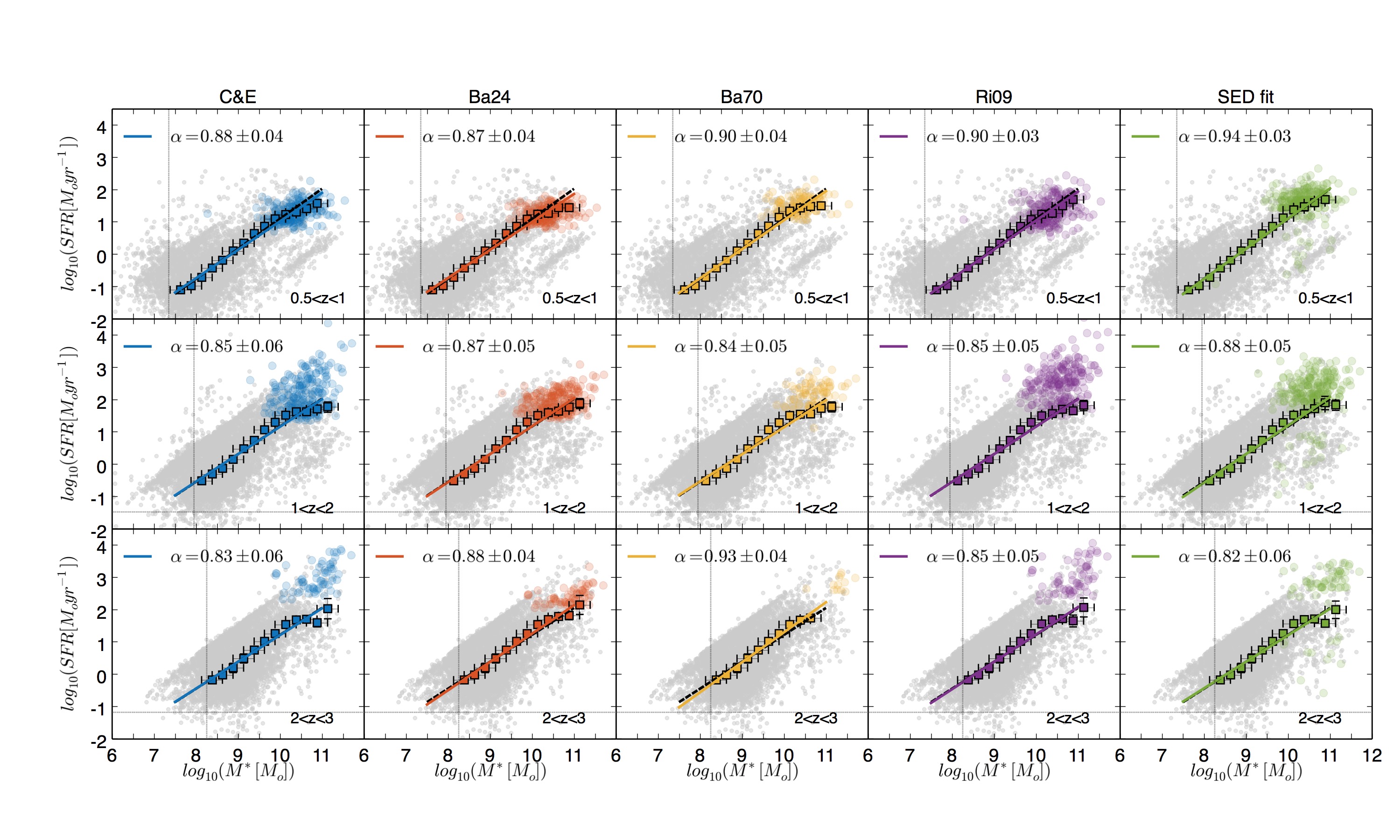}
 \caption{SFR versus stellar mass for different redshift bins. \textit{From top to bottom:} redshifts $0.5\leq z<1$, $1\leq z<2$ and $2<z<\leq3$. Grey points are all CANDELS data in the GOODS-S without IR observations and their SFR is derived from the UV. Coloured points in each panel are the $24 \, \rm \mu$m detected galaxies and in each column their SFR are obtained with different methods. \textit{From left to right:} C$\&$E, Ba24, Ba70, Ri09 and IR SED fitting. In each panel, the black dashed line shows the best fit for the MS for the CANDELS data with SFR derived only from the UV, as shown in Fig \ref{fig:MS_CANDELS}. The coloured lines show the best fit adding the $24 \, \rm \mu$m detected galaxies and the slope values of each MS is reported at the top left of each panel. Each MS is derived with an ODR fit using the Gaussian mean position derived fitting the sSFR distribution in each stellar mass bin. The mean positions of the MS Gaussian component in each stellar mass bin is shown in each panel with coloured squares. Error bars are the errors in the mean position of the MS Gaussian component derived from the fit.}
 \label{fig:SFR_IR_3}
 \end{figure*}

In Figure \ref{fig:SFR_IR_3}, we show the SFR-M$^{*}$ plane, with SFR$_{IR}$ derived using the five different L$_{IR}$ conversion formulae analysed in this paper. When IR observations are available we use the SFR$_{IR}+$SFR$_{UV}$, with UV SFR not corrected for dust extinction. On the other hand, when IR observations are not available, we use the dust-corrected, UV SFR. The MS is derived in each case by fitting an ODR to the Gaussian mean in the sSFR distribution in each 0.25 dex stellar mass bin. The best fit values derived using this method are listed in Table \ref{tab:MS_slope_IR_3}.\par

Observations in the IR are available only for galaxies with  log$_{10}(M^{*}/M_{o})\gtrsim$10, therefore in this work we can only analyse the impact of different L$_{IR}$ conversion formulae on the high-mass end of the MS and, in particular, on its flattening.  The MS slopes change by less than 0.11 with all methods and at all redshifts. In particular, at 0.5$\leqslant z<$1, the Ba24 method shows the flattest slope, 0.87$\pm$0.04, while the SED fit method has the steepest slope, 0.94$\pm$0.03. At 1$\leqslant z<$2, adding the IR observations decreases the slope by 0.01-0.02 dex for all methods, excluding the Ba24 and SED fit methods. At 2$\leqslant z <$3, the differences are slightly higher than in the lower redshift bin between -0.01 and +0.10 dex. However, taking into account the errors associated to each slope, all slopes derived by using only UV data or including IR observations are consistent with each other. \par

When including IR observations, the flattening of the MS changes depending on the L$_{IR}$ conversion formulae used and, in some cases, the MS is consistent with no flattening, even at the lowest redshift bin (i.e. Ri09). Therefore, a careful inclusion of IR observations is necessary to properly quantify the flattening of the MS at high stellar masses.\par

  \begin{table*}[htbp]
      \caption[]{Slope and intercept ($log_{10}SFR=a+b \cdot log_{10}(M^{*}$\-9)) of the main sequence in different redshift bins and for different SFR$_{IR}$ derivation methods. The MS is derived by using an ODR fitting to the mean position of the MS Gaussian component in the SFR distribution inside each 0.25 dex stellar mass bin.}
         \label{tab:MS_slope_IR_3}
     $$ 
     \bgroup
\def\arraystretch{1.5} 
    \begin{tabular}{l|cc|cc|cc}
        \hline
         Method  & \multicolumn{2}{|c|}{0.5$\leq z<1$} & \multicolumn{2}{|c|}{1$\leq z<2$} & \multicolumn{2}{|c}{2$\leq z<3$}\\
         \hline
         & a & b & a & b & a & b \\
        \hline
        SFR$_{UV}$ &0.18$\pm$0.03 & 0.93$\pm$0.03 & 0.32$\pm$0.06 & 0.86$\pm$0.05 & 0.40$\pm$0.05 & 0.83$\pm$0.05\\ 
        SFR$_{UV}$ $+$ SFR$_{IR}^{C\&E}$ &0.15$\pm$0.04 & 0.88$\pm$0.04 & 0.32$\pm$0.06 & 0.85$\pm$0.06 &0.39$\pm$0.06 & 0.83$\pm$0.06\\ 
        SFR$_{UV}$ $+$ SFR$_{IR}^{B24}$ & 0.14$\pm$0.04 & 0.87$\pm$0.04 & 0.32$\pm$0.05 & 0.87$\pm$0.05 & 0.39$\pm$0.04 & 0.88$\pm$0.04 \\ 
        SFR$_{UV}$ $+$ SFR$_{IR}^{B70}$ & 0.16$\pm$0.04 & 0.90$\pm$0.04 & 0.32$\pm$0.06 & 0.84$\pm$0.05 & 0.38$\pm$0.03 & 0.93$\pm$0.04 \\ 
        SFR$_{UV}$ $+$ SFR$_{IR}^{Ri09}$ & 0.15$\pm$0.03 & 0.89$\pm$0.03 & 0.32$\pm$0.06 & 0.85$\pm$0.05 & 0.39$\pm$0.05 & 0.85$\pm$0.05 \\ 
        SFR$_{UV}$ $+$ SFR$_{IR}^{fit}$ & 0.18$\pm$0.03 & 0.93$\pm$0.03 & 0.32$\pm$0.05 & 0.88$\pm$0.05 & 0.40$\pm$0.07 &0.82$\pm$0.07\\
        \hline
    \end{tabular} 
    \egroup 
     $$ 
   \end{table*} 
 \section{Summary and conclusions}\label{sec:summary}
In this work we studied the SFR-M$^{*}$ plane for a sample of 24463 galaxies at 0.5$\leqslant z\leqslant$3 with photometry from the UV through the near-IR. In particular, we firstly derived the MS with a linear fitting and, secondly, we studied quenched, star-forming, and starburst galaxies, separating them by fitting their sSFR with three Gaussian components. In addition, we analysed the impact of different bolometric IR luminosities conversion formulae on the MS of SF galaxies for a subsample with IR observations. \par

In our first method, we derived the MS of star-forming galaxies applying a method commonly used in the literature. Specifically, we made a linear regression for all SF galaxies, defined as all galaxies with log$_{10}(sSFR/yr^{-1})>$-9.8. The MS slope derived using this method depends on the considered stellar mass range, because of contamination from starbursts and quenched galaxies, whose fraction depends on the analysed stellar masses, and a possible flattening of the MS at high stellar masses. Therefore, it is important to carefully distinguish between quenched, star-forming, and starburst galaxies, in particular when considering a narrow stellar mass range, and, moreover, it is important to consider the same stellar mass range when comparing different works. Indeed, our results are consistent with the values found in the literature once we considered the same stellar mass range.  
\par

As a second step, we decomposed the sSFR distribution (in each stellar mass bin). Three components are evident (quenched, SF, and SB galaxies) with different importance depending on the stellar mass and the redshift considered. In particular, the SB population increases in number with increasing redshift, particularly for galaxies with log$_{10}(M^{*}/M_{o})<$9 between z$\sim$0.75, where starburst galaxies are less than 20$\%$, and z$\sim$1.5, where they are between 20 and 30$\%$, with respect to the total number of galaxies in the same stellar mass range. The SB fraction increases also with decreasing stellar mass. Overall, SB galaxies are $\sim5\%$ at z$=$0.5-1, $\sim12\%$ at z$=$1-2 and $\sim16\%$ at z$=$2-3, with respect to the total number of galaxies with log$_{10}(M^{*}/M_{o})=8.25-11.25$ of those redshifts. On the other hand, the quenched galaxies increase in number with stellar masses and they are a substantial number at log$_{10}(M^{*}/M_{o})>$10.5 at z$<$2 (20-60$\%$). The fraction of quenched galaxies also decreases with redshift. The slope of the MS changes from 0.93$\pm$0.03 at  $0.5\leq z<1$ to 0.83$\pm$0.05 at $2<z<\leq3$, decreasing with the redshift. However, a slight bending is present at high stellar masses (log$_{10}(M^{*}/M_{o})>$10.25) and the MS at lower masses is consistent with a slope of $\sim$ 1 and a mild evolution with redshift. A tight sequence of SB galaxies is also present, with a slope close to unity. On the other hand, quenched galaxies show a less clear sequence, they seem separated from the MS at log$_{10}(M^{*}/M_{o})>$10 and closer to the MS at lower stellar masses. \par

In addition, we analysed the impact of using different IR luminosity derivation methods on the high-mass end of the MS. In particular, we considered two methods to convert $24 \, \rm \mu$m luminosity to L$_{IR}$ (C$\&$E and Ba24), one to convert 70$\mu$m luminosity to L$_{IR}$ (Ba70), one to convert flux at $24 \, \rm \mu$m to SFR (Ri09), and a fifth one that derives the L$_{IR}$ by fitting 24-160$\mu$m observations with empirical IR SED templates. \par

In general, the slope of the MS changes by $<$0.11 when including the SFR$_{IR}$, with all different L$_{IR}$ derivation methods, both with a simple linear regression and decomposing the sSFR distribution with the Gaussian components. However, a proper analysis and inclusion of the SFR$_{IR}$ is essential when analysing the MS slope at high stellar mass and its possible flattening, which is indeed visible with UV-based SFR. Indeed, both the curvature and the starting stellar mass of the bending depend on the method used to derive the L$_{IR}$ and, remarkably, with some of the methods analysed in this work the MS is consistent with no flattening at high stellar masses even at the lowest redshift bin.\par

Overall, using different fitting techniques and different L$_{IR}$ derivation methods, we derived slopes between 0.87 and 0.93, at 0.5$\leqslant z<$1, between 0.84 and 0.87, at 1$\leqslant z<$2, and between 0.79 and 0.93 at 2$\leqslant z<$3. A dependence of the MS slope on the stellar mass is evident, therefore a flatter (steeper) MS slope could be derived limiting the study to high (low) stellar masses. Depending on the analysed redshift and stellar mass range, it could be necessary to distinguish between quenched, SF, and starburst galaxies in order to derive the MS without contaminants.\par
  
\section*{Acknowledgment}
This paper is based on observations taken by the CANDELS Multi-Cycle Treasury Program with NASA/ESA HST, which is operated by the Association of Universities for Research in Astronomy, Inc., under NASA contract NAS5-26555. \emph{Herschel} is an ESA space observatory with science instruments provided by European-led Principal Investigator consortia and with important participation from NASA. LB and KIC acknowledge the support of the Nederlandse Onderzoekschool voor de Astronomie (NOVA). KIC also acknowledges funding from the European Research Council through the award of the Consolidator Grant ID 681627-BUILDUP. We thank N.D. Padilla for useful discussions.

\bibliographystyle{aa}
\bibliography{mybib}

\onecolumn
\clearpage
 \begin{appendix} \section{Analysis of redshift outliers}\label{sec:outliers}
When comparing the photometric redshifts of the main sample used in this work with some spectroscopic redshifts presented in the literature, there are $\sim$11$\%$ of outliers with $\delta z=|z_{phot}-z_{spec}|/(1+z_{spec})>0.15$ (see Sect. \ref{sec:zphot_mass}). In this Appendix we analyse the distribution of these outliers in the SFR-M$^{*}$ plane and their effect on the general conclusions of the paper. \par
\begin{figure*}[]
\centering
  \includegraphics[width=0.9\linewidth, keepaspectratio]{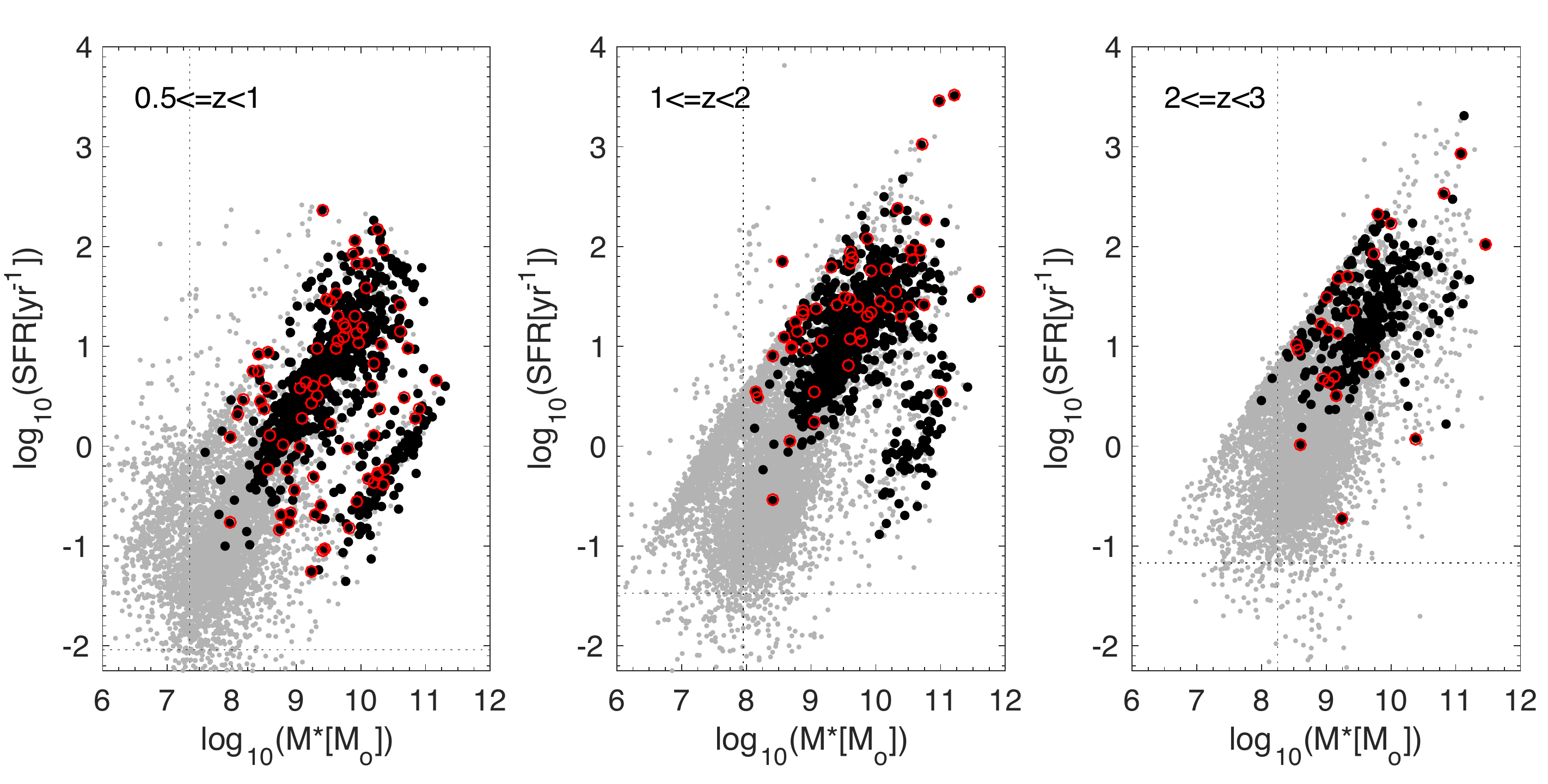}
 \caption{SFR-M$^{*}$ plane. Grey points are all galaxies in the sample analysed in this work, black dots are galaxies in the same sample with spectroscopic redshift, and red circles are galaxies that are outliers in redshift. Each column shows a different redshift bin: 0.5$\leqslant z<$1 (\textit{left}), 1$\leqslant z<$2 (\textit{central}) and 2$\leqslant z <$3 (\textit{right}). The vertical black dotted lines are the 90$\%$ completeness limit in stellar mass in each redshift bin. The horizontal dotted black lines are the 3$\sigma$ detection limit in SFR in each redshift bin.}
 \label{fig:fout_plane}
 \end{figure*}

Figure \ref{fig:fout_plane} shows the position of all galaxies with spectroscopic redshift and of all redshift outliers in the SFR-M$^{*}$ plane. Outliers are $8\pm1\%$ at redshift $0.5\leq z<1$ and $7\pm1\%$ at redshift $1\leq z<2$ and $2\leq z<3$, with no evident evolution in the analysed redshift range. From Figure \ref{fig:fout_plane} is evident that the redshift outliers do not systematically occupy a specific locus on the plane. Indeed, at $0.5\leq z<1$ outliers are distributed among starburst, MS galaxies and quenched galaxies. Only at $1\leq z<2$, it is evident that there are almost no outliers among the quenched galaxies, therefore the fraction of quenched galaxies at high stellar masses could be even higher than what has been derived in the paper. At the highest redshift bin the statistic starts to be poor and outliers are distributed among both SB and MS galaxies. \par
Overall, outliers do not occupy specific regions of the SFR-M$^{*}$ plane, therefore they do not qualitatively change the general conclusions of the paper. Only the fraction of QG with log$_{10}(M^{*}/M_{o})>$10 at $1\leq z<2$ may be underestimated. However, a more quantitative analysis of the effect of these outliers is not possible because galaxies with spectroscopic redshift are not representative of the full sample, they have generally log$_{10}(M^{*}/M_{\odot})\gtrsim9$; they are a minor fraction, only $\sim8\%$ of the CANDELS sample at $0.5<z\leq3$; and they do not have an homogeneous selection criteria.
 
  \section{Analysis of the need for a SB component}\label{sec:AIC}
\begin{figure*}[!htbp]
\centering
\includegraphics[width=1\linewidth, keepaspectratio]{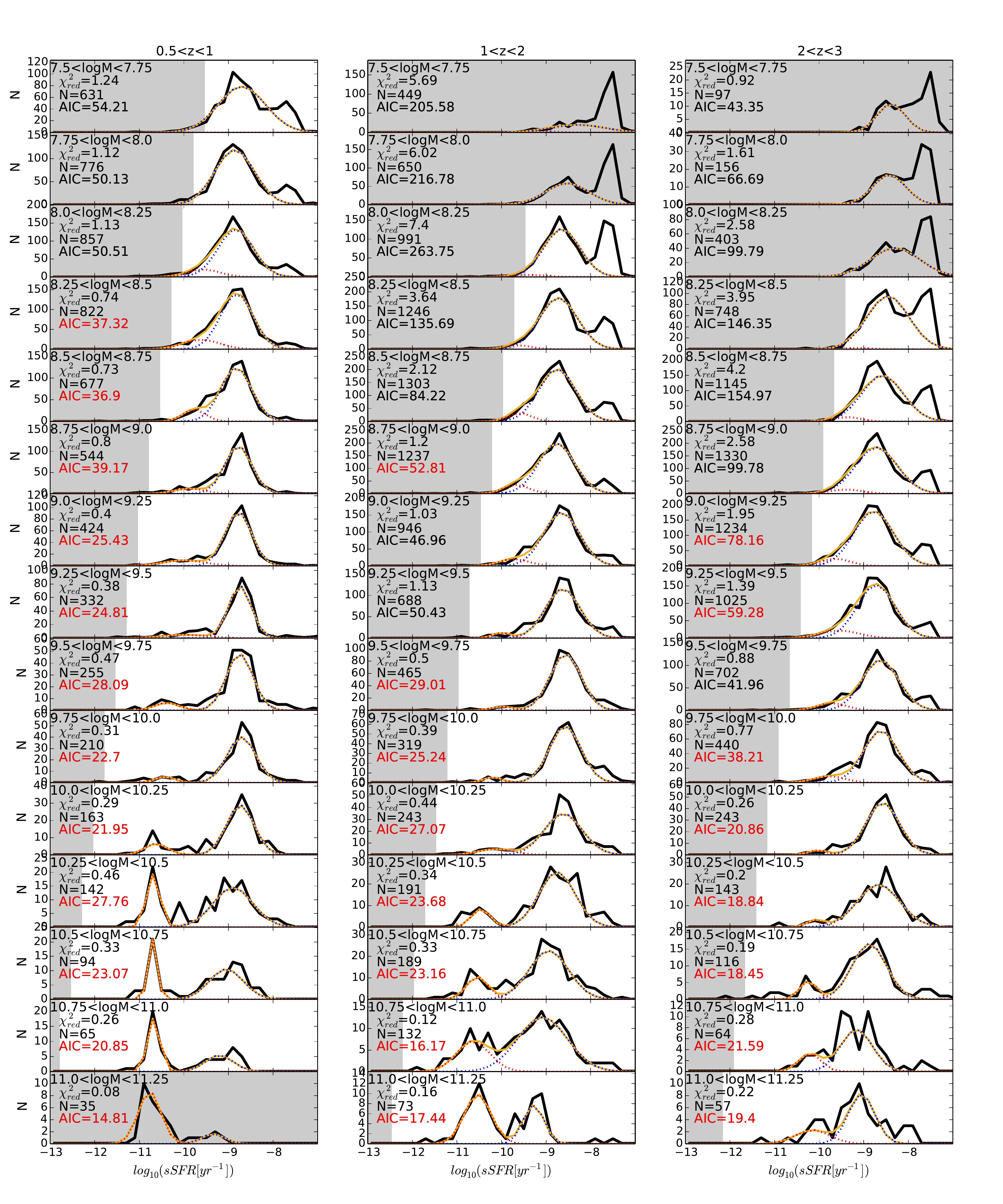}
\caption{sSFR distribution of the CANDELS sample in different z and stellar-mass bins. Each column shows a different redshift bin: 0.5$\leqslant z<$1 (\textit{left}), 1$\leqslant z<$2 (\textit{central}) and 2$\leqslant z <$3 (\textit{right}). Each row shows a different stellar mass bin, with a width of 0.25 dex, from 7.5$ \leqslant log_{10}(M^{*})<$7.75 (\textit{top}) to 11$ \leqslant log_{10}(M^{*})<$11.25  (\textit{bottom}). Black thick continuous lines show the sSFR distribution and the yellow thin continuous lines are the best fitted models. Dotted lines are the two different Gaussian components: quenched galaxies (\textit{red}) and MS galaxies (\textit{blue}). The $\chi_{red}^{2}$ and AIC values of each fit and the number of galaxies in each bin are shown in the top left of each panel. The AIC value coloured in red are smaller than the AIC values of the three-component fits in the same stellar mass and redshift bin (Fig. \ref{fig:Gauss0}). Grey areas show the sSFRs and stellar masses for which this sample is not complete (below the 3$\sigma$ limit in sSFR or below the 90$\%$ completeness in stellar mass), or has  statistics that are too low (N$<$50).}
\label{fig:Gauss0_2G}
 \end{figure*}

In this paper, we fit the sSFR distribution in each stellar mass and redshift bin using three Gaussian components. In this Appendix we repeat the fit using two Gaussian components, one for the MS galaxies and another for the quenched galaxies, in order to see whether the SB component is really necessary to describe the data. To compare the two fits we derive the Akaike Information Criterion\citep[AIC;][]{Akaike1973} defined as:
\begin{equation} \label{AIC}
AIC = -2\ln(\mathcal{L}) + 2k
,\end{equation}
where $\mathcal{L}$ is the likelihood function and k is the number of parameters in the model. This criterion takes into account the goodness of the fit and the complexity of the model at the same time. Among different models, the one that best represents the data with the minimum amount of parameters corresponds to the model with the smallest AIC value. We calculate the AIC value for both the three-component (Fig. \ref{fig:Gauss0}) and the two-component fits (Fig. \ref{fig:Gauss0_2G}) and we compare the AIC values in each stellar mass and redshift bin. \par
Unsurprisingly, the three-component fit is favoured by the data in the lowest-stellar-mass bins that also correspond to the stellar-mass bins where SB are more numerous. The stellar-mass bin limit below which the third component is necessary to fit the sSFR distribution increases with redshift from log$(M^{*}/M_{o})<$8.25 at $0.5\leq z<1$ to log$(M^{*}/M_{o})<$9 at $2\leq z<3$, following the SB fraction evolution with redshift. At high stellar masses galaxies are mainly in the MS or quenched, therefore a third component is not necessary. \par
As a summary, at low stellar masses and at high redshifts, SB galaxies are not negligible and a third component is necessary to properly describe the sSFR distribution.
 
\end{appendix}

\end{document}